\documentclass[
 amsmath,
 amssymb,
]{revtex4-2}

\pdfoutput=1 %arXiv

\usepackage{graphicx}% Include figure files

\usepackage{bm}% bold math
\usepackage{float}% table position

\begin{document}

\title{Estimating continuous data of wrist joint angles using ultrasound images}
\author{Yo Kobayashi}
\email{yo.kobayashi@es.osaka-u.ac.jp}%
\affiliation{Graduate School of Engineering Science, Osaka University, Osaka, Japan}%

\author{Yoshihiro Katagi}
\affiliation{Graduate School of Engineering Science, Osaka University, Osaka, Japan}%

% \date{\today}

\begin{abstract}
Ultrasound imaging has recently been introduced as a sensing interface for joint motion estimation. The use of ultrasound images as an estimation method is expected to improve the control performance of assistive devices and human--machine interfaces. This study aimed to estimate continuous wrist joint angles using ultrasound images. Specifically, in an experiment, joint angle information was obtained during extension--flexion movements, and ultrasound images of the associated muscles were acquired. Using the features obtained from ultrasound images, a multivariate linear regression model was used to estimate the joint angles. The coordinates of the feature points obtained using optical flow from the ultrasound images were used as explanatory variables of the multivariate linear regression model. The model was trained and tested for each trial by each participant to verify the estimation accuracy. The results show that the mean and standard deviation of the estimation accuracy for all trials were root mean square error (RMSE)=1.82 $\pm$ 0.54  deg and coefficient of determination (R2)=0.985 $\pm$ 0.009. Our method achieves a highly accurate estimation of joint angles compared with previous studies using other signals, such as surface electromyography, while the multivariate linear regression model is simple and both computational and model training costs are low.


\end{abstract}

\maketitle

\section{\label{sec:introduction}Introduction}

Recently, the aging of society has become a major problem in developed countries, where the population of elderly people aged 65 years and above is increasing, whereas that of the working population is decreasing. The number of cases in which frail elderly workers perform work is expected to increase owing to the aging of heavy industrial workers and the shortage of elderly caregivers. Therefore, methods to reduce the workload of the elderly and support their rehabilitation are urgently required. Solutions have been proposed, including upper- and lower-limb assistive devices such as motorized prostheses, exoskeleton robots, and soft exosuits. Numerous studies have been conducted on the motor support and rehabilitation of motor disabilities in humans. Prosthetic limbs are used to improve the quality of life of amputees by replacing the function and appearance of the lost upper or lower limbs. Research and development of prosthetic limbs with motors and other actuators are also being undertaken to improve functionality and operability \cite{clement2011bionic,tucker2015control}. Exoskeletal robots are expected to restore natural motor functions and improve quality of life by assisting the user's movement \cite{shi2019review, zhou2021lower, mcfarland2019considerations, islam2020exoskeletons}. Exoskeletal robots are wearable assistive devices with joint link mechanisms that transfer torque from actuators to human joints to reduce the workload and solve the problem of rehabilitation support. With the recent development of soft robotics, wearable assistive devices based on lightweight and flexible materials, such as cloth and silicone, called exosuits, have been developed \cite{xiloyannis2021soft, bardi2022upper, thalman2020review, ali2021systematic}. The exosuit fits better than exoskeletal robots and is lightweight and portable. Exosuits are used to support rehabilitation in the medical and rehabilitation fields and to reduce the workload in the life support field.

In such prostheses, exoskeletal robots, and exosuits, the user's movements must be measured or estimated to control their movements naturally \cite{tucker2015control, kato2018estimating, chiaradia2018design, dinh2017hierarchical, lotti2020adaptive}. More basically, advances in sensing, information processing, and mechanical devices have led to research on fluent and intuitive human--machine interfaces for the interaction of humans with digital systems, augmented/virtual reality interfaces, and physical robotic systems. The estimation of human joint motion is also critical for human--machine interfaces  that can respond accurately and quickly to user intentions \cite{zhao2020emg}. 

First, we describe a method for measuring the joint angles using a wearable sensor. When measuring the motion of the user, measurement is implemented with encoders or goniometers. These contact sensors are installed directly on the joints and thus provide highly accurate measurements. However, in many applications, placing a sensor without interfering with the integrity of the joint is difficult, and integrating encoders into wearable systems is uncomfortable and cumbersome \cite{woods2022joint}. Goniometers are also used for other single-joint motions \cite{edwards2004measuring}. However, goniometers require careful alignment with the center of the joint and must be repositioned each time the joint of interest changes, making it impossible to monitor multi-joint motion  \cite{bonnet2015monitoring}. In addition, both require avoiding interference in the mechanical arrangement of prosthetic limbs, exoskeletal robots, soft exosuits, and human–machine interface devices, which often become obstructions.

Next, we discuss the estimation of the joint angles using other sensor information. Surface electromyography (sEMG), which measures the electrical activity on the skin surface, is the primary biological signal used to estimate joint angles. sEMG uses noninvasive electrodes attached to the skin to measure the potential due to the excitation of motor units in the surface muscles \cite{zhang2021dual}. sEMG has long been used as a noninvasive human–machine interface to capture user intent and enable intuitive equipment control \cite{tucker2015control}. For example, Zhang et al. conducted a study to estimate continuous elbow joint angles from sEMG signals using a multiple-input autoregressive structure model \cite{zhang2013human}. Zhao et al. studied EMG-driven musculoskeletal models to estimate continuous wrist motion \cite{zhao2020emg}. Chen et al. estimated the joint angles of a human elbow online using a method called hierarchical projection regression, which focuses on learning high-dimensional sEMG data \cite{chen2013hierarchical}. Sawaguchi et al. used sEMG signals and a physical model to estimate the wrist joint angle \cite{Sawaguchi2011Wrist}. However, sEMG is susceptible to various factors, such as changes in electrode impedance, fatigue, sweat, and electrode misalignment \cite{tucker2015control}, and joint angle estimation with sEMG relies on signals that are difficult to maintain over time \cite{hargrove2013robotic}. Furthermore, muscle crosstalk makes accurately measuring the signal of the target muscle difficult owing to interference from other muscles in complex exercises such as multi-degree-of-freedom exercises \cite{raiteri2015ultrasound}.

Recently, ultrasound imaging has recently been introduced as a sensing interface for joint motion estimation \cite{zhang2021dual,guo2009performances}.
Ultrasound imaging has been proposed as an alternative noninvasive technique for joint motion estimation because of its high signal-to-noise ratio, direct visualization of the target tissue, and ability to access deep muscles \cite{zhang2021dual}.
Ultrasound images can detect deformations in both superficial and deep muscles and are sensitive to changes in muscle contraction \cite{guo2009performances}, enabling the direct measurement and differentiation of adjacent muscle contractions, even at different depths \cite{zhang2021dual}.
Thus, joint posture and motion estimation based on ultrasound images may not suffer from the limitations of sEMG \cite{zhang2021dual}. For example, Yang et al. conducted a study using A-mode ultrasound to perform finger joint motion classification \cite{yang2018towards}. Shi et al. performed finger flexion motion recognition using ultrasound images and processed them using a support vector machine \cite{shi2012recognition}. Sikdar et al. recognized individual finger movements from ultrasound images using the k-nearest-neighbor algorithm \cite{sikdar2013novel}. Huang et al. compared the accuracy of sEMG-based and ultrasound-based hand gesture classifications \cite{huang2017ultrasound}. MacIntosh et al. conducted a study to classify finger gestures from ultrasound images using multilayer perceptrons and support vector machines \cite{mcintosh2017echoflex}. Recently, research has been conducted on the use of information obtained from ultrasound images to control exosuits. For example, Nuckols et al. used ultrasound images to estimate the soleus muscle in real time and used this information to control an exosuit \cite{nuckols2020automated}. They also used ultrasound imaging to estimate the force generated by the soleus muscle and controlled the exosuit to be proportional to the estimated force \cite{nuckols2021individualization}.

Currently, sEMG is primarily used as a biological signal in control methods for assistive devices such as prosthetic limbs, exoskeletal robots, exosuits and human-machine interface devices. The use of ultrasound images as an alternative estimation method is expected to improve the performance of assistive devices. However, previous research on motion recognition using ultrasound images has focused on the classification of motions and gestures. If joint angles can be continuously estimated using ultrasound images, ultrasound images can be more useful for the control of assistive devices and human--machine interface devices.

This study aimed to estimate continuous wrist joint angles using ultrasound images. Specifically, in the experiment, joint angle information was obtained during extension--flexion movements, and ultrasound images of the associated muscles were acquired. Using the features obtained from ultrasound images, a multivariate linear regression model was used to estimate the joint angles. The coordinates of the feature points obtained using optical flow from the ultrasound images were used as explanatory variables of the multivariate linear regression model. The model was trained and tested for each trial by each participant to validate the estimation accuracy.

\section{\label{methods} Methods}

This study was approved by the Ethics Committee for Human Studies at the Graduate School of Engineering Science, Osaka University. All experiments were performed according to the approved design. All the participants signed an informed consent form before starting the experiment. The participants included in the figures provided informed consent for the publication of the images in an online open-access publication. This study was conducted in accordance with the principles of the Declaration of Helsinki.

\begin{figure*}
\includegraphics[width=18cm]{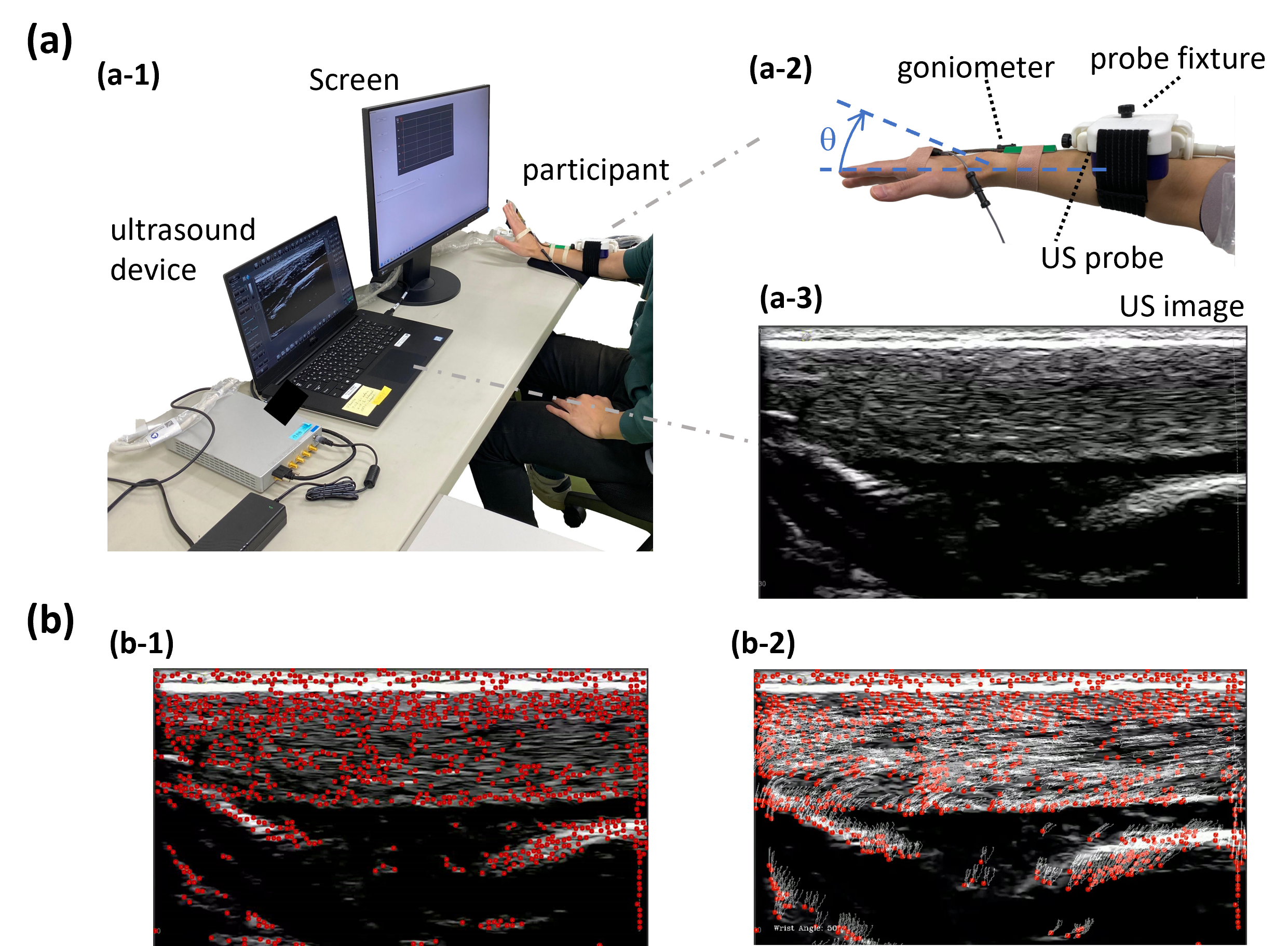}% Here is how to import EPS art
\caption{\label{fig:concept}
Experimental setup and image processing. \textbf{(a)} Photograph of the experimental setup employed in this study. The wrist joint angle (a-2) and ultrasound images (a-3) of the muscles associated with the movement in the flexion-extension direction of the wrist joint were acquired from human participants (a-1). \textbf{(b)} To determine the initial positions of the feature points, we used the Shi--Tomasi corner detection method for the first frame of the ultrasound images. The red dots in (b-1) are an example of initial positions of feature points obtained using Shi--Tomasi corner detection. The coordinates of each feature point extracted by corner detection were tracked using optical flow to obtain coordinate data. (b-2) shows the example of the optical flow image after several cycles of wrist motion. Here, the series of gray dots indicates the path of the coordinates.
}
\end{figure*}

Data were collected from 10 healthy young participants aged 20 years or older who had completed their growth spurt. The test method was the same for all participants. The experimental setup is shown in Fig. \ref{fig:concept}. A goniometer (SG110/A, Biometrics Ltd.) was attached to the wrist of each participant to measure the wrist joint angle. An ultrasound probe (LF11-5H60-A3, Telemed Medical Systems) of an ultrasound imaging system (ArtUS EXT-1H, Telemed Medical Systems)  was placed  using a specific band (ProbeFix Dynamic T, Telemed Medical Systems) near the elbow joint, where large muscle deformations were observed during flexion and extension of the wrist joint. The ultrasound probe was placed at a position (approximately 50 mm away from the elbow joint in the direction of the wrist joint) where the extensor carpi radialis shortus, a muscle that contributes greatly to wrist flexion and extension, and its surrounding muscles could be observed. The direction of the ultrasound probe was determined to be longitudinal to the forearm.

The participants sat naturally with the right forearm supported by an armrest on a table and the palm facing downward. The participants performed wrist extension exercises targeting the red curve (target joint angle) displayed in the experimental measurement application (Fig. \ref{fig:concept}). The target joint angle was a repeating sine wave with 0 deg as the reference position (Fig. \ref{fig:concept}). For the amplitude of each sine curve, 20, 40, or 60 deg was randomly selected (ref. Fig. \ref{fig:time}). The period of the target joint angle was set to 4 s. In each experiment, 18 cycles of specified movements were measured. Owing to the tendency for noise to occur at the beginning and end of each experiment, 15 cycles of data were used for the analysis (the first two cycles and the last cycle of each dataset were removed). The data used in the validation phase included the joint angles of all amplitudes (20, 40, and 60 deg) (ref. Fig. \ref{fig:time}). The joint angles and ultrasound images were measured during the experiments. Five sets of experiments were performed per participant, resulting in 50 sets for 10 participants.

Time-series data of the wrist joint angles were acquired at 1000 Hz using a goniometer. These data were resampled at 63 Hz to match the frame rates of the ultrasound images. A low-pass filter (Butterworth filter \cite{butterworth1930theory}) was applied to the data to remove noise. The data were standardized (Z-score normalization) to perform ridge regression. These data were used as input vectors for the joint angles \textit{$ \boldsymbol\theta= \left (\theta_{1}, \theta_{2}, \cdots   \right )$}, where \textit{$\theta_{k}$} is the filtered and standardized joint angle at the k-th time data point.

Ultrasound images of the muscle during flexion and extension of the wrist joint were acquired at 63 Hz, and the feature points in the ultrasound images were extracted by image processing at each time point. To determine the initial positions of the feature points, we used the Shi--Tomasi corner detection method \cite{shi1994good} for the first frame of the ultrasonic images. The coordinates of each feature point extracted by corner detection were tracked using optical flow \cite{lucas1981iterative} to obtain coordinate data \textit{$ \mathbf{\bar{s}} = \left (\bar{x}^{1},\bar{y}^{1}, \bar{x}^{2}, \bar{y}^{2}, \cdots   \right ) $}, where (\textit{$\bar{x}^{i}$}, \textit{$\bar{y}^{i}$} ) is the coordinate of the i-th feature point. Here, only the coordinates of the feature points that could be tracked to the end by the optical flow were used. The average number of feature points used for the explanatory variable vector was 995. Next, standardization (Z-score normalization)  and a low-pass filter (Butterworth filter) were applied to the time series data for each coordinate \textit{$ \mathbf{\bar{s}}^i = (\bar{s}^{i}_1,\bar{s}^{i}_2, \cdots)$} to obtain filtered and standardized feature point vector \textit{$ \mathbf{s}^i = (s^{i}_1,s^{i}_2, \cdots)$} at i-th feature point. From these data, the matrix \textit{$\mathbf{S} = (\mathbf{s_{1}},\mathbf{s_{2}}, \cdots)$} of explanatory variables for multivariate regression analysis was calculated, where \textit{$ \mathbf{s_k} = \left (s^{1}_k,s^{2}_k, s^{3}_k, s^{4}_k, \cdots  \right )^T$} is the vector of explanatory variables at each time point k. 

A multivariate linear regression model was then used to estimate the joint angles from the coordinates of the feature points in the ultrasound images as follows. Note that the following description relates to the method of estimating one joint angle or posture in one experiment for a single experimental dataset. 

Equation (\ref{eq:1}) was used to calculate the estimated joint angle $\hat{\theta}_k$ at the k-th time data point in the multivariate linear regression model.

\begin{eqnarray}
\hat{\theta}_k = \sum_{i} w^i  s^i_k
\label{eq:1}
\end{eqnarray}

where \textit{$\hat{\theta}_{k}$} is the estimated joint angle at the k-th time data point, \textit{$s_k^{i}$} represents the coordinates of the feature point of the i-th location at the k-th time data point, and \textit{$w^i$} is the weight corresponding to the coordinates  at the i-th feature point.

The multivariate linear regression model was trained using ridge regression, which is a method for calculating the weight \textit{$\mathbf{w}=(w^{1},w^{2}, \cdots)$}, minimizing the loss function (\ref{eq:a1}) with the residual sum of the squares term and L2 regularization term. Ridge regression was used to avoid overlearning and multicollinearity. Lasso regression and elastic net regression were also tested; however, their accuracies were similar to or lower than those of ridge regression. Unlike ridge regression, lasso and elastic net regression cannot be optimized using analytical expressions. Ridge regression was chosen because it may be preferable from the standpoints of simplicity and speed.

\begin{eqnarray}
L=\sum_{k}\left(\theta_k-\hat{\theta}_k\right)^2+\frac{1}{2} \lambda \sum_{i} (w^{i})^2
\label{eq:a1}
\end{eqnarray}
where \textit{$\theta_{k}$} is the measured joint angle at each time data point k.

The weight \textit{$\mathbf{w}$} that minimizes the loss function was determined analytically using the following equation.

\begin{eqnarray}
\mathbf{w}= \boldsymbol \theta  \mathbf{S}^{\mathrm{T}} \left( \mathbf{S} \mathbf{S}^{\mathrm{T}}+\lambda \mathbf{I}\right)^{-1}
\label{eq:a2}
\end{eqnarray}

where 
\textit{$\mathbf{w}=(w^{1},w^{2}, \cdots)$} is the weight vector; 
\textit{$\boldsymbol \theta = \left (\theta_{1}, \theta_{2}, \cdots   \right )$} is the joint vector summarizing the measured joint angle at each time data point;
\textit{$\mathbf{S}= (\mathbf{s^{1}},\mathbf{s^{2}}, \cdots)$} is the  matrix of explanatory variables  summarizing the vectors of the measured cordinate data;
\textit{$ \mathbf{s^i} = \left (s^{i}_1,s^{i}_2, \cdots   \right )^T $} is the coordinate vector of the i-th feature point;
\textit{$\lambda $} is a hyperparameter and \textit{$\mathbf{I}$} is the identity matrix.
The hyperparameter  \textit{$\lambda $} was determined by trial and error, and the same value (\textit{$\lambda=10$}) was used for all the training sessions.

Using the above method, the estimated joint angles \textit{$\hat{\theta}_{k}$} were obtained in all trials. Learning and estimation were performed for each trial in each dataset and a different learner was used for each trial. Four-fifths of the total data were used as training data to train the model, and the remaining one-fifth were used as validation data to evaluate the estimation accuracy. For each dataset, the measured and estimated joint angles were subjected to an inverse transformation of standardization and returned to their original scale. The root mean square error (RMSE) and coefficient of determination (R2) between the estimated and measured joint angles in the validation data were calculated to evaluate the accuracy of this method.

\section{\label{result} Result}

\begin{figure*}
\includegraphics[width=18cm]{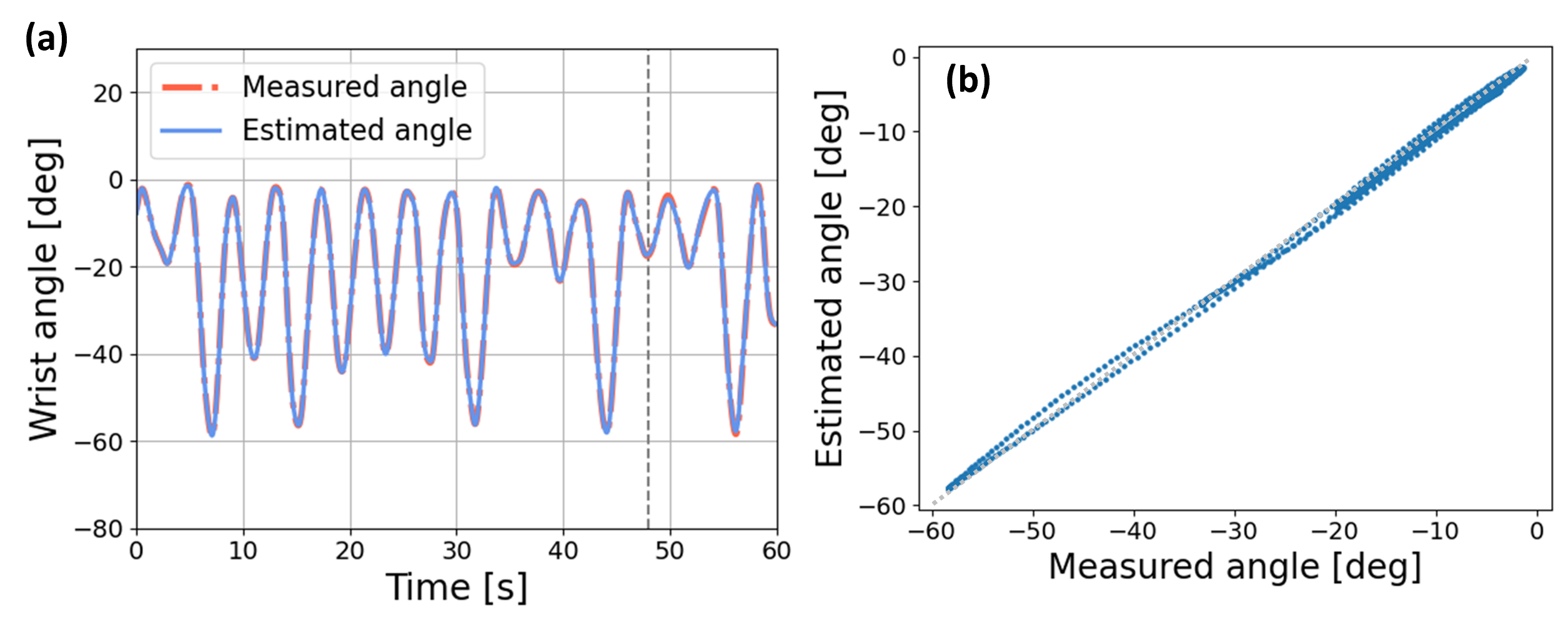}% Here is how to import EPS art
\caption{\label{fig:time} 
Representative examples of wrist joint angle estimation results. \textbf{(a)}  The red line is the measured value and the blue line is the estimated value. The left side of the vertical gray dotted line is the training data and the right side is the validation data. The RMSE and R2 of this trial were 0.855 deg and 0.997, respectively.  \textbf{(b)} Relationship between measured and estimated wrist joint angles in the validation data section. The gray dotted line is the y=x line. The measured and estimated values are on the y=x line, and there is no large hysteresis, so the measured and estimated values are in good agreement. These results results indicate that a highly accurate estimation was achieved in estimating joint angles using ultrasound image.} %f3
\end{figure*}

A representative example of the time-series data of the wrist joint angle estimation results is shown in Fig \ref{fig:time}(a). The mean and standard deviation of the estimation accuracy for all trials were RMSE=1.82 $\pm$ 0.54 deg and R2=0.985 $\pm$ 0.009. Figure \ref{fig:time}(b) shows the relationship between the measured and estimated wrist joint angles in the validation phase of the trial shown in Fig. \ref{fig:time}(a). The results show that the measured and estimated values lie on the y=x line (gray dotted line), and no significant hysteresis occurred; thus, the measured and estimated values agree well. These results indicate that a highly accurate estimation was achieved by estimating the joint angles and posture using ultrasound images.

\begin{figure*}
\includegraphics[width=9cm]{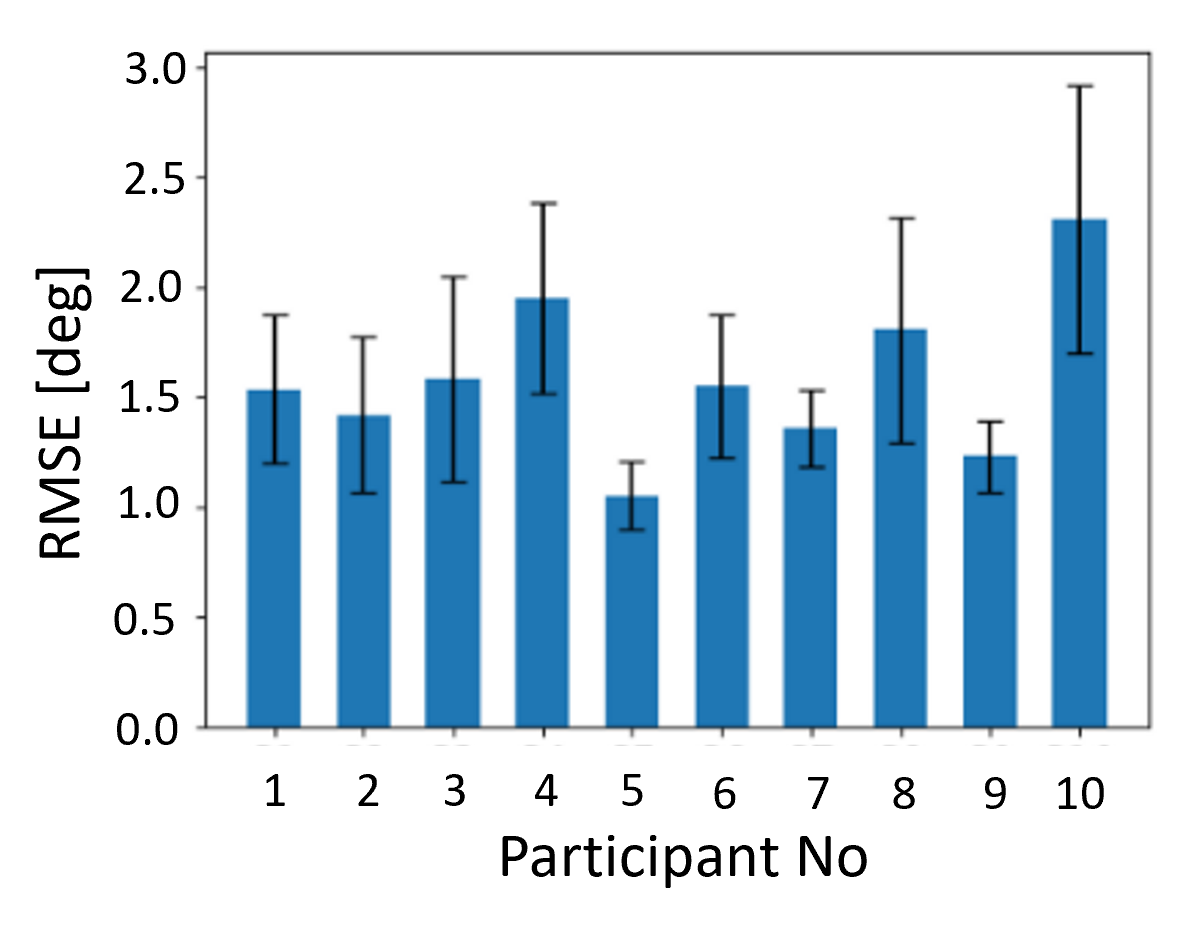}% Here is how to import EPS art
\caption{\label{fig:participant} 
RSME results for estimation accuracy for each participant. From this results, we can conclude that the accuracy of the estimation did not decrease substantially depending on the participant.
} %f3
\end{figure*}

The RSME results for the estimation accuracy for each participant are summarized in Fig. \ref{fig:participant}. In addition, the RSME and R2 of the estimation accuracy for each participant are summarized in Table \ref{result_by_participant_table}. For Participant No. 5, with the highest accuracy (No. 5), the mean and standard deviation of the RMSE were 1.05 $\pm$ 0.15  deg, and the mean and standard deviation of R2 were 0.995 $\pm$ 0.001. For the participant with the lowest accuracy (No. 10), the mean and standard deviation of the RMSE were 2.31 $\pm$ 0.61 deg, and the mean and standard deviation of R2 were 0.979 $\pm$ 0.010. From these results, it can be concluded that the accuracy of the estimation did not decrease substantially depending on the participant.

\begin{table}[H]
    \centering
    \caption{RMSE and $R^2$ of each participant.}
    \label{result_by_participant_table}
    \begin{tabular}{|c|c|c|c|c|}
        \hline
         & RMSE(mean)[deg] & RMSE(std)[deg] & $R^2$(mean) & $R^2$(std)  \\ \hline\hline
        No. 1 & 1.66 & 0.24 & 0.990 & 0.003 \\ \hline
        No. 2 & 1.95 & 0.72 & 0.983 & 0.012 \\ \hline
        No. 3 & 1.50 & 0.59 & 0.988 & 0.009 \\ \hline
        No. 4 & 2.17 & 0.61 & 0.980 & 0.009 \\ \hline
        No. 5 & 1.15 & 0.20 & 0.994 & 0.003 \\ \hline
        No. 6 & 2.05 & 0.30 & 0.983 & 0.003 \\ \hline
        No. 7 & 1.80 & 0.73 & 0.988 & 0.009 \\ \hline
        No. 8 & 2.45 & 1.15 & 0.971 & 0.028 \\ \hline
        No. 9 & 1.41 & 0.14 & 0.991 & 0.002 \\ \hline
        No. 10 & 2.07 & 0.68 & 0.983 & 0.011 \\ \hline
    \end{tabular}
\end{table}

\section{\label{discussion} Discussion}

The contribution of this study is that the wrist joint can be measured with high accuracy using ultrasound images. Quantitatively, the proposed method achieved a highly accurate estimation of the wrist joint angle with an RMSE $\fallingdotseq$1.8 deg and R2$\fallingdotseq$0.99. The average accuracies of previous studies that have continuously estimated wrist or elbow joint angles are listed below: In the study by Zhao et al., wrist joint angles were estimated using an EMG-driven musculoskeletal model, and the accuracy was RSME$\fallingdotseq$10 deg, R$\fallingdotseq$0.9 \cite{zhao2020emg}. The accuracy was RSME$\fallingdotseq$5.0 deg in a study by Sawaguchi et al., in which the wrist joint angle was estimated from the sEMG signal using a physical model \cite{Sawaguchi2011Wrist}. In a study by Zhang et al., in which elbow joint angles were estimated from sEMG signals, the accuracy was RSME$\fallingdotseq$8.3 deg \cite{zhang2013human}. In a study by Kato et al., in which the wrist joint angles were estimated from skin deformation near the muscle using a multivariate linear regression model, the accuracy was RSME$\fallingdotseq$5.4 deg \cite{kato2018estimating}. Thus, our method achieves a highly accurate estimation of the joint angles compared with previous studies. However, it should be noted that simple comparisons may not be appropriate due to different experimental conditions.

\begin{figure*}
\includegraphics[width=18cm]{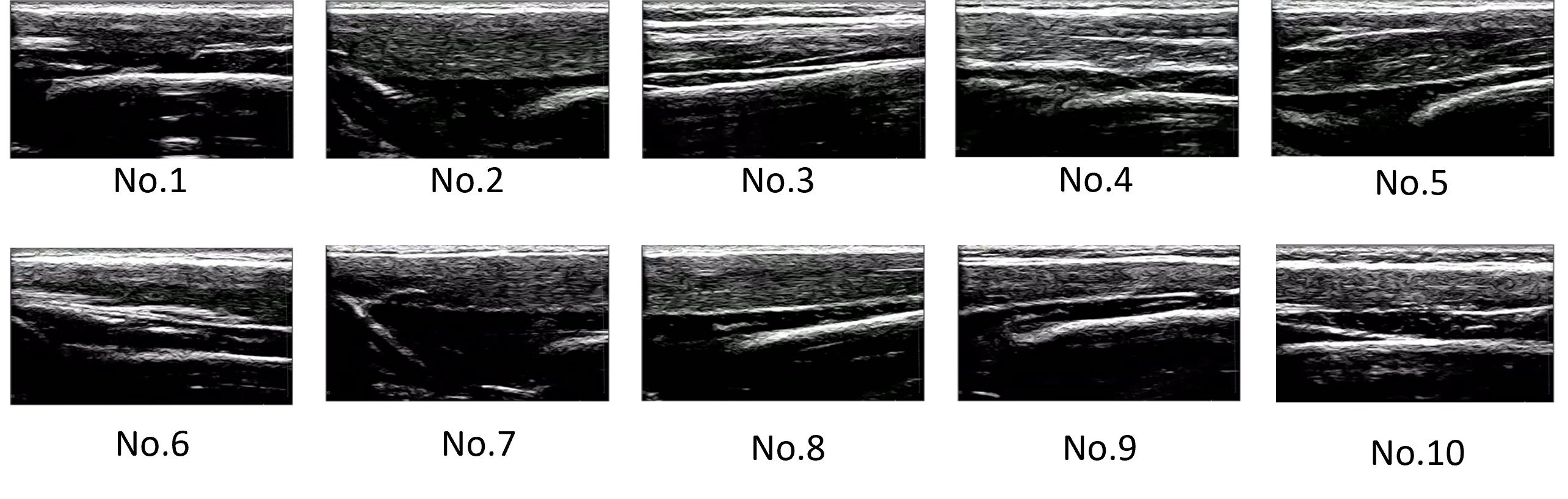}% Here is how to import EPS art
\caption{\label{fig:image} 
The ultrasound images acquired during the experiment of each participant. The anatomical structure of the forearm, such as the thickness of the muscles, differs slightly from participant to participant, and the position of the ultrasound probe during the measurement differs slightly from participant to participant.
} %f3
\end{figure*}

In this study, the mean estimation error had an accuracy of 1.8 deg, and no participant deviated substantially from this. However, the estimation errors differed for each participant. One of the factors contributing to this difference among the participants was the difference in the ultrasound images acquired during the experiment (Fig. \ref{fig:image}). This may be due to the fact that the anatomical structure of the forearm, such as the thickness of the muscles, differs slightly from participant to participant, and the position of the ultrasound probe during the measurement differs slightly from participant to participant. In this study, the ultrasound probe was attached to the outer part of the wrist, approximately 5 cm from the elbow joint in the direction of the wrist joint. However, it was difficult to adjust the position so that similar ultrasound images were obtained in all participants only by visual inspection or palpation at the time of attachment. Additionally, the position and direction of the ultrasound probe (the section of the forearm that was acquired as the ultrasound image) could not be examined in this study. As the forearm muscles move significantly in the longitudinal direction during wrist motion, it was thought that muscle motion could be properly tracked by cutting parallel to the forearm muscles. However, optimal placement of the ultrasound probe must also be considered. By changing the position and direction of the ultrasound probe placement and comparing the estimation accuracy, it is possible to further improve the accuracy by determining the optimal placement method. However, the results of this study showed that it was possible to estimate the wrist joint angle with the comparable accuracy for all participants. Therefore, the results indicated that it was possible to estimate the wrist joint angles around the ultrasound probe attachment position, which was the target of this study, regardless of the definite position of the ultrasound probe.

In this study, a multivariate linear regression model was used to estimate the wrist joint angles, and the model was trained and estimated for each trial. This method can be used with small amounts of data, and both computational and model training costs are lower than those of complex machine learning methods, such as neural networks and deep learning. This helps to create models tailored to individual users. However, compared with complex machine learning methods, there are limitations in the application of training models, such as the fact that a model trained on one participant cannot be used by another.

Although the probe used in this study is one of the smallest commercially available probes, it is still large and rigid, which may hinder its use in wearable scenarios. In the future, as this method will be used in assistive devices and human–machine interface devices, it will be necessary to make ultrasound probes smaller, more flexible, and wearable. Currently, progress has been made in the research and development of ultrasound probes and transducers to make them smaller, more flexible, and wearable \cite{hu2018stretchable, pashaei2019flexible, elloian2022flexible, wang2022bioadhesive,hu2023stretchable, noda2023textile}. We take advantage of this progress by using ultrasound imaging in a manner that does not interfere with the wearer and is easy to use within the system.

In recent years, research on physical reservoir computing, in which the dynamics of soft materials are regarded as information-processing devices such as recurrent neural networks, has suggested the possibility of using various physical phenomena as computational resources \cite{tanaka2019recent, nakajima2020physical}. In physical reservoir computing, various states (explanatory variables) in a physical nonlinear system are used, and information is processed by extracting these states using learned linear and static readouts  (often a multivariate linear regression model). These studies investigated the computational capabilities of the musculoskeletal system \cite{sumioka2011computation} using the soft spine as a physical reservoir to generate the motion of a quadruped robot  \cite{zhao2012embodiment}, processed information using a soft body such as an octopus  \cite{nakajima2015information, nakajima2014exploiting, nakajima2013soft, nakajima2018exploiting}, and used soft material bodies as computational resources. Another study proposed a soft wearable suit with tactile sensors, which can be considered as a tactile sensor network for monitoring natural body dynamics and as a computational resource for estimating human and robot postures \cite{sumioka2021wearable}. In the proposed method, the deformation inside the muscle was used as an information processing device based on the concept of physical reservoir computing, and the human muscle was used as a computational resource for joint angle estimation. In other words, the dynamics of the muscle are regarded as a physically implemented recurrent neural network, and the wrist joint angle of the participant is estimated from the internal deformation using only linear and static readouts (multivariate linear regression model), such as physical reservoir computing. It uses human muscles as a computational resource and can, therefore, process information at very low learning and computational costs. We name these types of research area ``biomechanical reservoir computing," which utilizes the physical dynamics occurring in the human (organisms) interior and in contact with surrounding environment as a computational resource, and are promoting researches such as \cite{kobayashi2023information} using this framework.

\section*{data availability}

The datasets generated in this study are available from the corresponding author upon request. The data are not publicly available to protect the privacy of the participants.

\begin{acknowledgments}
This work was supported in part by a Grant-in-Aid for Scientific Research from the Ministry of Education, Culture, Sports, Science, and Technology (MEXT) (19H02112, 19K22878 and 23K18479) of Japan. This work was supported in part by JST, PRESTO, Grant Number JPMJPR23P2 (Japan).
\end{acknowledgments}

\bibliographystyle{unsrt}

\bibliography{Manuscript} % Produces the bibliography via BibTeX.

\end{document}